\documentclass[onecolumn,
 amsmath,amssymb,
 aps,
nofootinbib
]{revtex4-1}

\usepackage{graphicx}
\usepackage{dcolumn}
\usepackage{xcolor}
\usepackage[section]{placeins}

\usepackage[hyperfootnotes=true]{hyperref}
\hypersetup{colorlinks, citecolor=blue, linkcolor=red, urlcolor=purple}

\begin{document}

\preprint{APS/123-QED}

\title{Bubble bursting jets are driven by the purely inertial collapse of gas cavities}

\author{Jos\'e M. Gordillo}
\email{jgordill@us.es}
\author{Francisco J. Blanco--Rodr\'{i}guez}%
\affiliation{\'Area de Mec\'anica de Fluidos, Departamento de Ingenier\'ia Aeroespacial y Mec\'anica de Fluidos, Universidad de Sevilla, Avenida de los Descubrimientos s/n 41092, Sevilla, Spain
}%

\date{\today}

\begin{abstract}
The analysis of numerical simulations describing the collapse of capillary cavities reveals that the jets originated from the bursting of bubbles are driven by the condition that the dimensionless liquid flow rate per unit length directed towards the axis of symmetry, $q_\infty$, remains nearly constant in time. This observation, which is justified in physical terms because liquid inertia prevents appreciable changes in $q_\infty$ during the short time scale characterizing the jet ejection process, together with the fact that bubble bursting jets are produced from the bottom of a conical cavity, justify the purely inertial scalings for the jet width and velocity found here, $r_{jet}\propto\sqrt{q_\infty\tau}$ and $v_{jet}\propto \sqrt{q_\infty/\tau}$, with $\tau$ indicating the dimensionless time after the jet is ejected, a result which notably differs from the common belief that the jet width and velocity follow the inertio-capillary scaling $r_{jet}\propto \tau^{2/3}$ and $v_{jet}\propto \tau^{-1/3}$. Our description reproduces the time evolution of the jet width and velocity for over three decades in time, obtaining good agreement with numerical simulations from the instant of jet inception until the jet width is comparable to that of the initial bubble. 
\end{abstract}

\maketitle

\begin{figure*}
	\centering
	\includegraphics[width=0.9\textwidth]{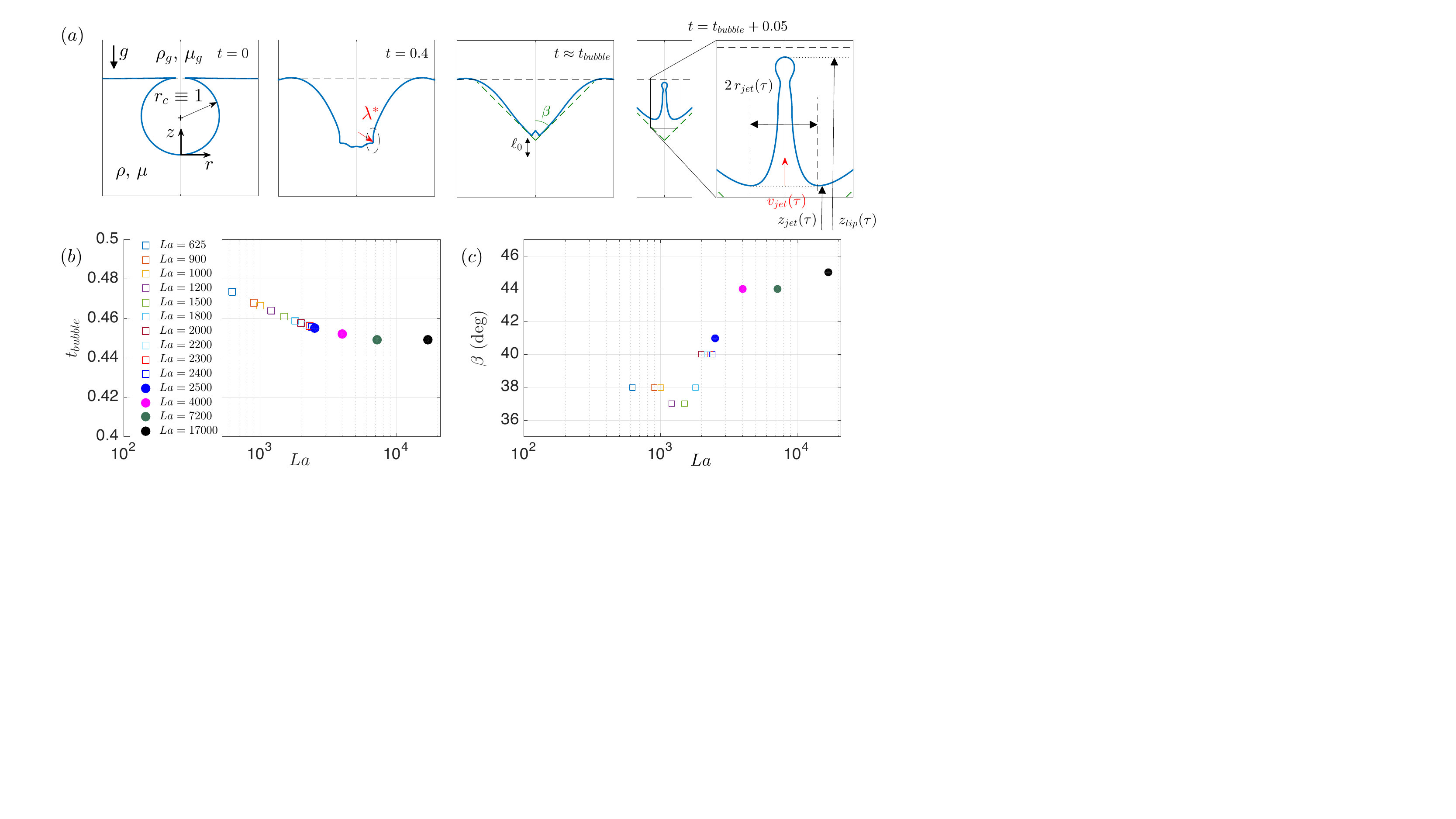}
	\caption{(a) Time evolution of the bubble collapse and subsequent jet ejection processes corresponding to a value of the Laplace number $La=Oh^{-2}=7200$. The numerical simulations have been carried out for a value of the Bond number $Bo=0.01$, as we did in \citep{JFM2020}, where we provide details on the numerical implementation, see also \cite{PRF2023} for numerical details. The last panel in $(a)$ shows the base of the jet, which is the point at the interface $r=r_s(z,\tau)$ of coordinates $r=r_{jet}(\tau)$, $z=z_{jet}(\tau)$ where $\partial r_s/\partial z=0$. The jet velocity is defined as $v_{jet}(\tau)=v_z(r=0,z=z_{jet})$, see also \cite{PRF2023}.(b) Values of $t_{bubble}$, with $t_{bubble}$ the instant of time at which the capillary waves with a wavelength $\lambda^*\propto Oh^{1/2}$ reach the bottom of the cavity. Since capillary waves propagate with a velocity which is independent of $La$ \cite{JFM2019}, $t_{bubble}$ is almost independent of $La$. (c) The value of the opening semiangle $\beta$ depends on $La$: for $La\geq 2500$ a bubble is not entrapped beneath the bottom of the cavity and $\beta=45^\circ$; for $La<2500$, a satellite bubble is entrapped and $\beta<45^\circ$, see Figs. \ref{fig1SM}-\ref{fig3SM}.} 
   \label{fig1}
\end{figure*}

Due to their widespread presence in nature and in technological applications, an increasing number of numerical \citep{Duchemin,PRL2009,Deike,Berny}, experimental \citep{Ghabache2,Thoroddsen_18,Thoroddsen_2020} and theoretical \citep{Zeff,PRLEggers} studies have been devoted to characterize the width and speed of the jets ejected after the collapse of a cavity or a bubble, a process which is ubiquitously present in our daily life experience. A type of jets exhaustively studied in recent times because of their crucial role played in the generation of nanometric sea spray aerosol (SSA) are the jets produced after the bursting of bubbles see e.g., \cite{MacIntyre,Bigg,AnRevVeron,PRLGanan,CommentBursting,PNAS,Matar,NatureFeng,AnnRevDeike,PNASVillermaux,PRLBird,JieFengNatPhys}. However, in spite of the many contributions on the subject, there does not exist a consensus in the community neither on the precise physical mechanism driving this type of jets nor on the equations describing the radius and the velocity of the first drop ejected. Then, it will be our purpose here and also in the companion paper \cite{PRF2023} to present a self-consistent description of the dynamics governing the time evolution of bubble bursting jets making use of scaling arguments, theory and of numerical simulations of the type depicted in Fig. \ref{fig1}, which have been carried out using \texttt{GERRIS} \citep{Popinet2003,Popinet2009} because, as it has been shown in \cite{Deike,JFM2020}, this numerical code has been proven to accurately reproduce experimental results related with the bursting of bubbles.

 In the cases relevant for sea spray aerosol production, the initial radii $R_b$ of the bubbles is such that the Bond number verifies the condition $Bo=\rho\,g\,R^2_b/\sigma\ll 1$, with $g$ indicating gravity, $\rho$ the liquid density and $\sigma$ the interfacial tension coefficient and, therefore, the jet ejection process is controlled by three dimensionless parameters namely, the Ohnesorge number, defined as $Oh=\mu/\sqrt{\rho\,R_b\,\sigma}$, with $\mu$ indicating the liquid dynamic viscosity and also by the density and viscosity ratios, respectively defined as $\rho_g/\rho=1.2\times 10^{-3}$ and $\mu_g/\mu=1.8\times 10^{-2}$, with $\rho_g$ and $\mu_g$ the gas density and viscosity, see Fig. \ref{fig1}. Since, in this study, the values of $\mu_g/\mu$ and $\rho_g/\rho$ are kept constant, the results reported here depend only on $Oh$ or, equivalently, on the Laplace number $La=Oh^{-2}$. Notice that all the variables appearing in the text have been made dimensionless using as characteristic values of length, time and pressure $R_b$, $R_b/V_\sigma$ and $\rho V^2_\sigma$ respectively, with $V_\sigma=\sqrt{\sigma/(\rho R_b)}$ indicating the capillary velocity. The origin of times, $t=0$, is set at the instant when the interface of the initially spherical bubble starts deforming, $t_{bubble}$ denotes the instant when the capillary wave reaches the bottom of the cavity and $\tau=t-t_{bubble}$ indicates instants of time after the capillary waves have reached the bottom of the cavity, see Fig. \ref{fig1}, being our main purpose here to deduce equations for the jet width $r_{jet}(\tau)$, for the vertical position of the base of the jet, $z_{jet}(\tau)$, and for the jet velocity $v_{jet}(\tau)$, see the rightmost panel in Fig. \ref{fig1}(a). In what follows, we report the analysis of the numerical results obtained for these three time-dependent variables for values of the Laplace number within the range $625 \leq La\leq 17000$, with each value of $La$ identified using the colour code of Fig. \ref{fig1}(b). 

The jets depicted in Fig. \ref{fig1}(a) are emitted once the capillary wave with a dimensionless wavelength $\lambda^*\propto Oh^{1/2}$, propagating with the $Oh$-independent dimensionless speed $\approx 5$, reaches the bottom of the cavity at the instant $t_{bubble}$, see Fig. \ref{fig1}(b) and \citep{JFM2019}. The capillary wave deforms the initial spherical bubble into a truncated conical surface with a half-opening angle $\beta(La)$ such that $\beta\simeq 45^\circ$ for $La\geq 2500$, see the green dashed line in the third panel of Fig. \ref{fig1}(a), Fig. \ref{fig1}(c) and also Figs. \ref{fig1SM}-\ref{fig3SM}. For $La\geq 2500$, the minimum radius of the truncated cone is given by \citep{JFM2019,JFM2020}:
\begin{equation}
r_{jet0}=0.2215\left(1-\sqrt{\frac{Oh}{0.0305}}\right)\, ,\label{rjet0}
\end{equation}
and, for our subsequent purposes, it proves convenient to define here $r^*_{jet0}=r_{jet0}(La=2500)\simeq 0.05$. It is shown in Figs. \ref{fig1SM}-\ref{fig3SM} that the topology of the interface changes for $La<2500$ since, in these cases, a tiny bubble is entrapped beneath the cavity before the jet is ejected. For $La<2500$, the jet is also issued from the base of a truncated conical surface with a  radius $r_{jet0}(La<2500)<r^*_{jet0}$. and with an opening semiangle which verifies $\beta(La<2500)<45^\circ$, see Fig. \ref{fig1}(c).

\begin{figure}
	\centering
	\includegraphics[width=0.45\textwidth]{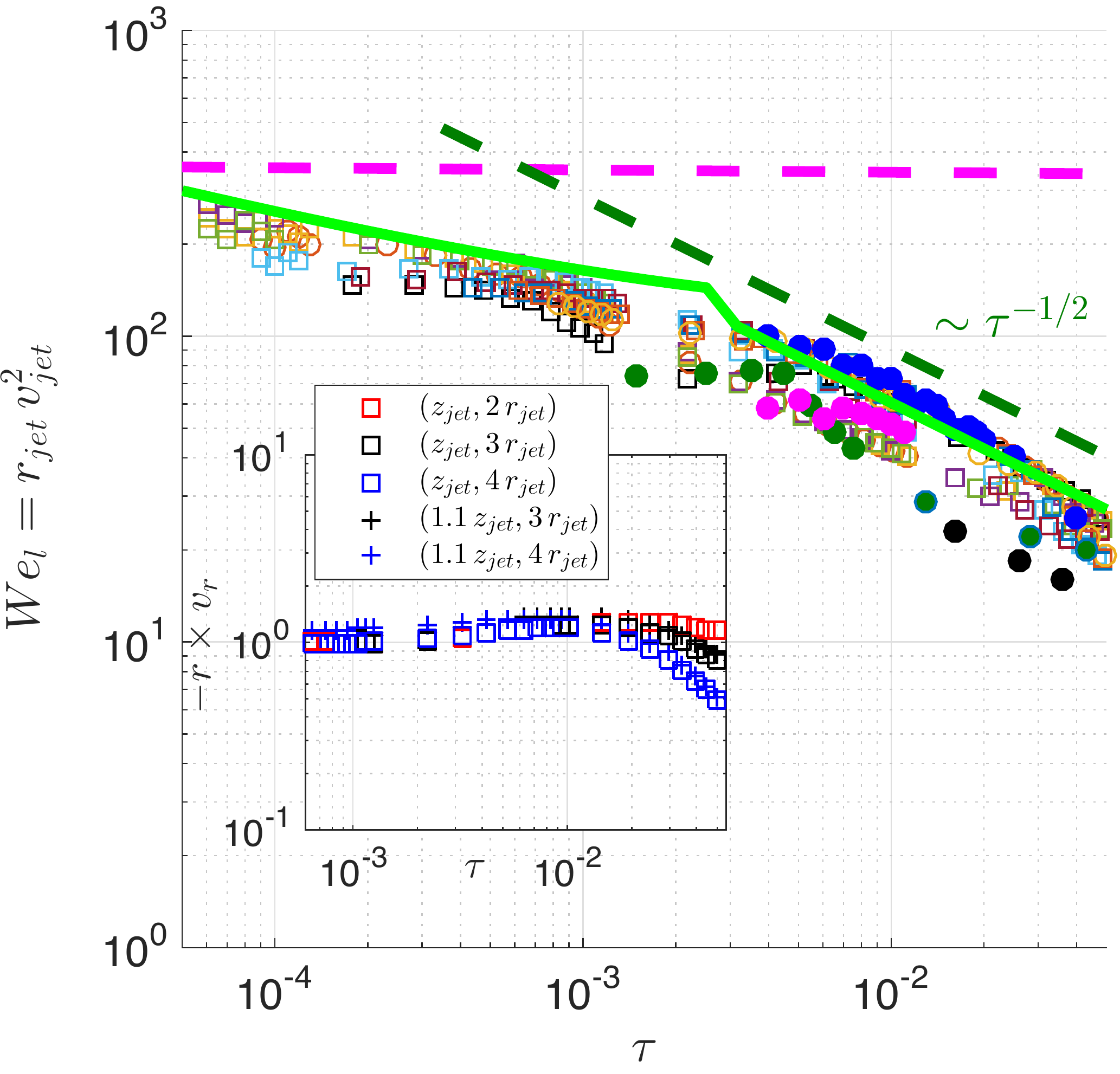}
	\caption{The value of the local Weber number is such that $We_l=r_{jet} v^2_{jet}\gg 1$, with $We_l\propto \tau^{-1/2}$ for $\tau\gtrsim 10^{-3}$, a result indicating that $r_{jet}\propto \sqrt{q_\infty \tau}$ and $v_{jet}\propto \sqrt{q_\infty/\tau}$, as it is explained in the text using the results in the inset, where it is shown that the values of the flow rate per unit length calculated using \texttt{GERRIS} as  $q=-r v_r$ for $La=2400$ and different values of $r/r_{jet}(\tau)$ and $z/z_{jet}(\tau)$, remain nearly constant along two decades in time. The green line indicates the predicted values of $We_l(r_{jet}(\tau))=v^2_{jet}\,r_{jet}=\left[3.4\,q_\infty(r_{jet}(\tau))\right]^2/r_{jet}(\tau)$, with $r_{jet}$ and $v_{jet}$ calculated using Eqs. (\ref{ecsBB})-(\ref{qinftyrjet}).} 
   \label{fig3}
\end{figure}
\begin{figure}
	\centering
	\includegraphics[width=0.4\textwidth]{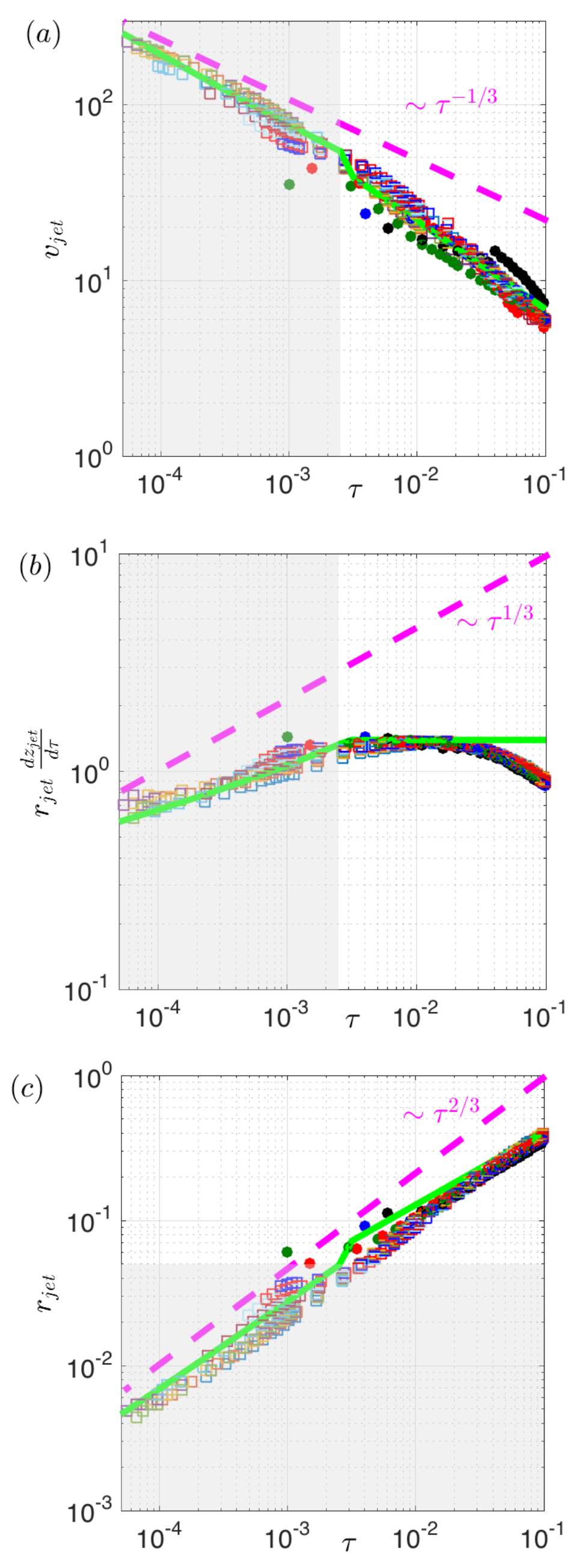}
	\caption{The values of ($a$) $v_{jet}(\tau)$, ($b$) $r_{jet} dz_{jet}/d\tau$ and ($c$) $r_{jet}(\tau)$ calculated using Eqs. (\ref{ecsBB})-(\ref{qinftyrjet}) and plotted using green lines, reproduce fairly well the numerical results. In contrast, the inertio-capillary predictions, plotted using pink lines, $v_{jet}\propto \tau^{-1/3}$, $r_{jet} dz_{jet}/d\tau\propto \tau^{1/3}$ and $r_{jet}\propto \tau^{2/3}$ substantially deviate from the numerical results. The shaded area indicates conditions corresponding to $La<2500$ for which $q_\infty$ is given by the second equation in (\ref{qinftyrjet}). The values in the figure correspond to $\tau>\tau^*(La)$, with $\tau^*(La)>0$ the instant of time such that $\left(z_{tip}-z_{jet}\right)/r_{jet}\geq 1.25$, see Fig. \ref{fig1}(a) and Fig. \ref{fig5SM}.} 
   \label{fig4}
\end{figure}

While the results reported in \cite{JFM2019,JFM2021} are limited to analyze right the instant of time at which the jet is ejected, \cite{PRLEggers} studied the spatio-temporal evolution of the jet shapes, finding that their results are compatible with the inertio-capillary scaling $r_{jet}(\tau)\propto \tau^{2/3}$; in contrast, we find here that this scaling is not valid to quantify the instants \emph{after} the jet is ejected. Indeed, the inertio-capillary scaling is characterized by the fact that the capillary and the dynamic pressure terms in the governing equations balance each other and, if the conclusions in \cite{PRLEggers} were correct, the value of the dimensionless number expressing the ratio between the dynamic and capillary pressures namely, the local Weber number defined as $We_l=v^2_{jet} r_{jet}$, should remain constant in time along the whole jet ejection process. However, the results depicted in Fig. \ref{fig3} reveal that: i) $We_l\gg 1$ for all values of $\tau>0$, a fact already suggesting that the jet dynamics cannot be driven by capillary forces and ii) $We_l$ is far from being a constant since $We_l\propto \tau^{-1/2}$ for $\tau\gtrsim 10^{-3}$.

The reason why $We_l=v^2_{jet} r_{jet}\propto \tau^{-1/2}$ for $\tau\gtrsim 10^{-3}$ is explained in the inset of Fig. \ref{fig3}, where we find that the value of the dimensionless flow rate per unit length $q=-r\,v_r$ remains approximately constant in time for different values of $r/r_{jet}(\tau)$ and $z/z_{jet}(\tau)$, with $v_r$ denoting the radial velocity. Indeed, if $r_{jet}\propto z_{jet}\propto \tau^{\alpha}$, velocities are proportional to $v_{jet}\propto \tau^{\alpha-1}$ and, since the results shown in the inset of Fig. \ref{fig3} indicate that $r_{jet} v_{jet}\propto \tau^{2\alpha-1}\approx const$, we conclude that $\alpha=1/2$, which explains the result $We_l=v^2_{jet} r_{jet}\propto \tau^{-1/2}$ for $\tau\gtrsim 10^{-3}$ depicted in Fig. \ref{fig3}. In physical terms, the reason why the radial flow rate per unit length directed towards the axis of symmetry remains approximately constant in time is due to the fact that liquid inertia ensures that this quantity does not change appreciably during the very short instants of time following the emergence of the jet \cite{PRL2009,JFM2010,JFM2021}. Consequently, the physical mechanism driving the ejection of bubble bursting jets can be summarized as follows: during the capillary collapse of the cavity, the liquid is accelerated inwards, inducing a radial inflow per unit length towards the axis of symmetry which, in dimensionless terms, is $q_\infty\approx 1$. Then, along the very short transient $\tau\ll 1$ during which the conical cavity collapses -see Figs. \ref{fig1}-\ref{fig3}, the flow rate $q_\infty$ acts as the far field boundary condition driving the jet ejection process, see also our theory in the companion paper \cite{PRF2023}. The constancy of the flow rate per unit length forces the inward motion of the conical cavity walls -see Fig. \ref{fig1}- and, therefore, the jet is issued as a mere consequence of mass conservation \cite{JFM2019,JFM2021}. In \cite{PRF2023} we quantify the ejection of jets from collapsing cavities with arbitrary shapes and, among other findings, we also report that the jets emanating from the collapse of conical surfaces for a constant value of $q_\infty$, are self-similar: in fact, the results depicted in Fig. \ref{fig4SM} reveal that, indeed, the time varying shapes of bubble bursting jets overlap onto the the purely inertial self-similar function we have calculated in \cite{PRF2023} using the results of the numerical code in \cite{Reuter}, employed by those authors to model the jets ejected by cavitation bubbles in very close proximity to a wall. 

Figure \ref{fig4} provides another clear evidence supporting that bubble bursting jets emerge as a consequence of the fact that $q_\infty$ remains constant during the short time interval $\tau\ll 1$ along which the jet ejection process takes place. Indeed, the equations deduced in \cite{PRF2023}, 
\begin{equation}
\begin{split}
v_{jet}=2\frac{dz_{jet}}{d\tau}=\frac{1.5K(\beta)}{\tan\beta}\sqrt{\frac{q_\infty(r_{jet})}{\tau}}\, , \quad r_{jet} v_{jet}\simeq 3.4\,q_\infty\, \label{ecsBB}
    \end{split}
\end{equation}
with $K(\beta\approx 40^\circ)\approx 1.6$ and $\beta(La)$ given in Fig. \ref{fig1}(c), closely follow the numerical results corresponding to $625 \leq La\leq 17000$. In contrast, the analogous scalings deduced using the inertio-capillary ansatz, $r_{jet}\propto \tau^{2/3}$, $v_{jet}\propto \tau^{-1/3}$ and  $r_{jet} d z_{jet}/d\tau\propto \tau^{1/3}$ clearly deviate from the data. The results in Fig. \ref{fig4} have been obtained for the slowly-varying values of the forcing flow rate per unit length $q_\infty$ given by
\begin{equation}
\begin{split}
&\mathrm{If}\quad r_{jet}\geq r^*_{jet0}=0.05: \quad q_{\infty}=0.82\\ &
\mathrm{If}\quad r_{jet}<r^*_{jet0}=0.05:\quad q_\infty(r_{jet})=\frac{-r_0\dot{r}_0(0)}{\left[r_0/\left(2r_c\right)(0)\right]}\times\\&\times \exp\left(-\sqrt{\left[\ln\left(\frac{r_0(0)}{2r_c(0)}\right)\right]^2-\ln\left(\frac{r_{jet}}{r_0(0)}\right)^2}\right),\label{qinftyrjet}
\end{split}
\end{equation}
see Fig. \ref{fig6SM}, where we find that $r_0(0)=0.2$, $r_0\,\dot{r}_0(0)=-1.5$ and $r_0(0)/(2\,r_c(0))=0.16$ for the whole range of $La<2500$ considered here. The reason for the different expressions for the far field forcing $q_\infty$ in Eq. (\ref{qinftyrjet}) can be explained as follows: for $r_{jet}>r^*_{jet0}\simeq 0.05$, the radius of the jet is sufficiently large for the flow rate per unit length $q_\infty\approx 1$ driving the jet ejection process to be fixed by the radial flow induced by the capillary collapse of the cavity at the length scale of the original bubble. However, for the cases corresponding to $La<2500$, $r_{jet}(\tau\ll 1)\ll 1$ and, in addition, in these cases a tiny satellite bubble is entrapped beneath the bursting bubble. The satellite bubbles depicted in Figs.  \ref{fig1SM}-\ref{fig3SM} and  \ref{fig6SM} are produced as a result of the purely inertial collapse of a cylindrical gas thread and, consequently, when $r_{jet}\lesssim 0.05$, the flow rate per unit length $q_\infty<0.82$ forcing the ejection of the jet is fixed by the value of the \emph{local} flow rate per unit length driving the purely inertial pinch-off of the bubble at the instant when the collapsing bubble radius coincides with that of the jet. Hence, the result in Eq. (\ref{qinftyrjet}) for the case $r_{jet}<0.05$ simply reflects the fact that the flow rate driving the pinch-off of bubbles \cite{PRLBurton2005,PRL2005,PRLGiant,JFM2006,PRL2007,PoF2008a}, see also the Appendix \ref{AppA} for details, is a decreasing function of the bubble radius.

With regard to the initial values of the jet width and velocity, for the cases $La\geq 2500$, the equation for $r_{jet0}$ is nothing but Eq. (\ref{rjet0}) whereas the initial jet velocity follows from Eq. (\ref{ecsBB}) as $v_{jet0}\simeq 2.5/r_{jet0}$ \cite{JFM2021}.
For the cases in which a bubble is entrapped namely, $La<2500$, a jet will be ejected following the solution of Eqs. (\ref{ecsBB})-(\ref{qinftyrjet}) whenever viscous effects are negligible namely, whenever the local Reynolds number $Re_l=Oh^{-1} r_{jet} v_{jet}\gtrsim O(1)\Rightarrow q_\infty\geq K\,Oh$ \cite{JFM2019}, with $K$ a constant. Therefore, using the result in Eq. (\ref{qinftyrjet}), the equations for the initial jet width and velocity deduced from $q_\infty=K\,Oh$ when a bubble is entrapped, read:
\begin{equation}
\begin{split}
&r_{jet0}=r_0(0)\times\\&\exp\left(-\frac{1}{2}\left[\left[\ln\left(\frac{-K Oh\,r_0(0)}{\left(2r_c r_0\dot{r}_0\right)(0)}\right)\right]^2-\left[\ln\left(\frac{r_0(0)}{2r_c(0)}\right)\right]^2\right]\right)\\ &
v_{jet0}=\frac{3.4 q_{\infty}(r_{jet0})}{r_{jet0}}=\frac{3.4 K Oh}{r_{jet0}}\label{r0v0entrapment}
\end{split}
\end{equation}
where we have made use of the second of the equations in (\ref{ecsBB}). The predictions in Eqs. (\ref{r0v0entrapment}) taking $K=10$ are close to the numerical values depicted in Fig. \ref{fig7SM}\footnote{The scalings deduced \emph{in the case} the inertio-capillary balance held right before the instant of jet ejection imply $q_\infty\propto r^{1/2}_{jet0}$ and, from the condition for jet ejection when a bubble is entrapped that the local Reynolds number is of order unity i.e., $q_\infty\propto Oh$, it follows that $r_{jet0}\propto Oh^2$ and, from the second of the equations in (\ref{r0v0entrapment}), $v_{jet0}\propto Oh^{-1}$, recovering the results in \cite{JFM2019}. The reason why the result of solving Eqs. (\ref{r0v0entrapment}) is similar to the one deduced assuming that the inertio-capillary scaling holds when the jet is ejected \cite{JFM2019} can be understood in view of the results depicted in Fig. \ref{fig4}, which reveal that the exponents for $r_{jet}(\tau)$ and $v_{jet}(\tau)$ become very close to $2/3$ and $-1/3$ respectively as $\tau\rightarrow 0$, this being an effect already explained in \cite{PRLGiant,JFM2021}, where it was shown that the exponents describing the purely inertial pinch-off of bubbles become similar to the ones corresponding to an inertio-capillary collapse when the initial aspect ratio of the cavity is not large, as it is the case here.} Anyway, the values of $r_{jet0}$ and $v_{jet0}$ deduced above when a bubble is entrapped are not the ones of the initial drop radius and velocity which, however, can be expressed as a function of $r_{jet}(\tau)$ and of $v_{jet}(\tau)$ following the results in \cite{JFM2020}, where the ballistic equations deduced in \cite{JFM2010} are coupled with the mass and the momentum balances at the top drop when this latter balance incorporates the relative flux of momentum, the capillary retraction term and also the drag force exerted by the gas. 

To conclude, here we have shown that the jets emerging from the collapse of millimetric or even micron-sized bursting bubbles are driven by a purely inertial mechanism in which the far field flow rate per unit length forces the collapse of the cavity walls and, hence, the emergence of a jet is a consequence of mass conservation, a picture that conceptually differs from the classical inertio-capillary balance proposed in \cite{PRLEggers}. 

 
\begin{acknowledgements} 
This work has been supported by the Grant PID2020-115655G, financed by the Spanish MCIN/ AEI/10.13039/501100011033.\\

\emph{Author contributions:} JMG designed the research and the theory, analyzed the data and wrote the paper, FJBR performed the numerical simulations and analyzed the data. All authors reviewed the results and approved the final version of the manuscript.
\end{acknowledgements}

\newpage

\appendix*
\section{Supplementary material}
\label{AppA}
\subsection{Bubble entrapment and value of $\beta$ at $t=t_{bubble}$.}

\begin{figure}
	\centering
	\includegraphics[width=\textwidth]{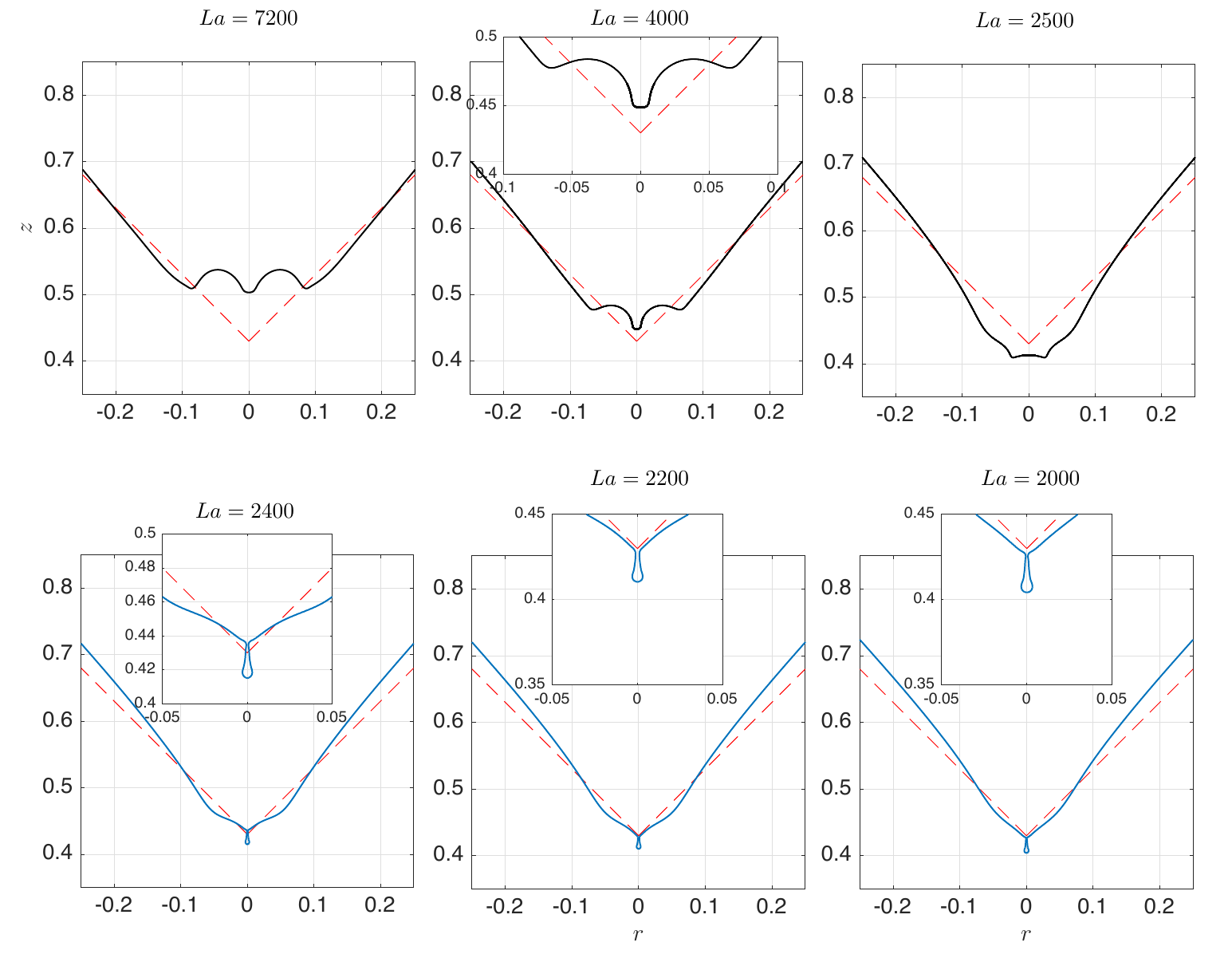}
	\caption{} 
   \label{fig1SM}
\end{figure}

\begin{figure}
	\centering
	\includegraphics[width=\textwidth]{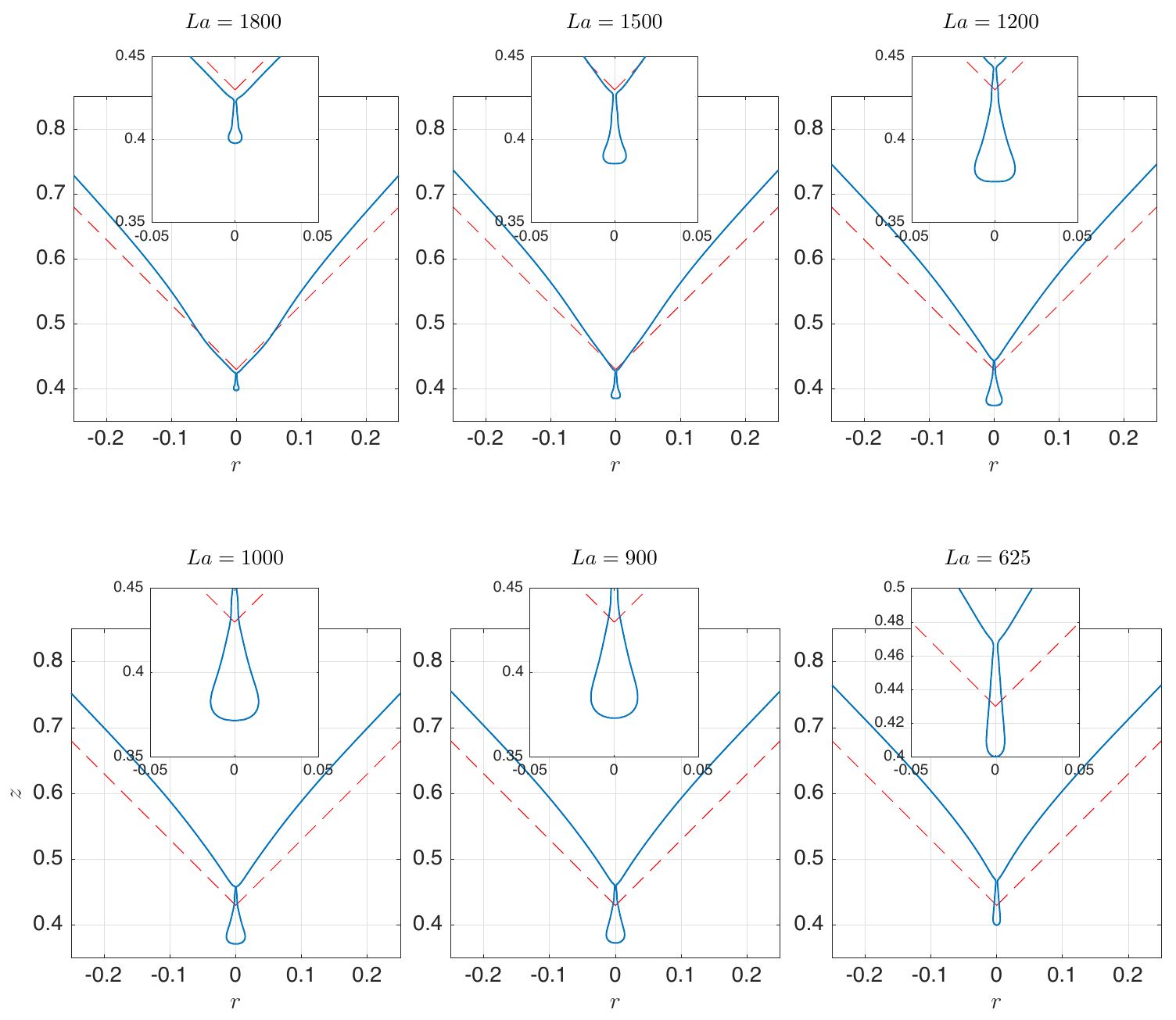}
	\caption{} 
   \label{fig2SM}
\end{figure}

Figures \ref{fig1SM}-\ref{fig2SM} show the local shape of the cavity at the instant $t_{bubble}$ the capillary waves reach the bottom of the bubble. Notice that, for $La\geq 2500$, a tiny satellite bubble is not entrapped beneath the cavity and the semiangle of the truncated conical surface from which the jet is ejected is very close to $\beta=45^\circ$, with this angle being indicated in the figure using red dashed lines. In contrast, for $625\lesssim La<2500$, a satellite bubble is entrapped and $\beta\simeq 37^\circ$ for $La\approx 1000$, see also Fig. 1(c) in the main text.

\begin{figure}
	\centering
	\includegraphics[width=\textwidth]{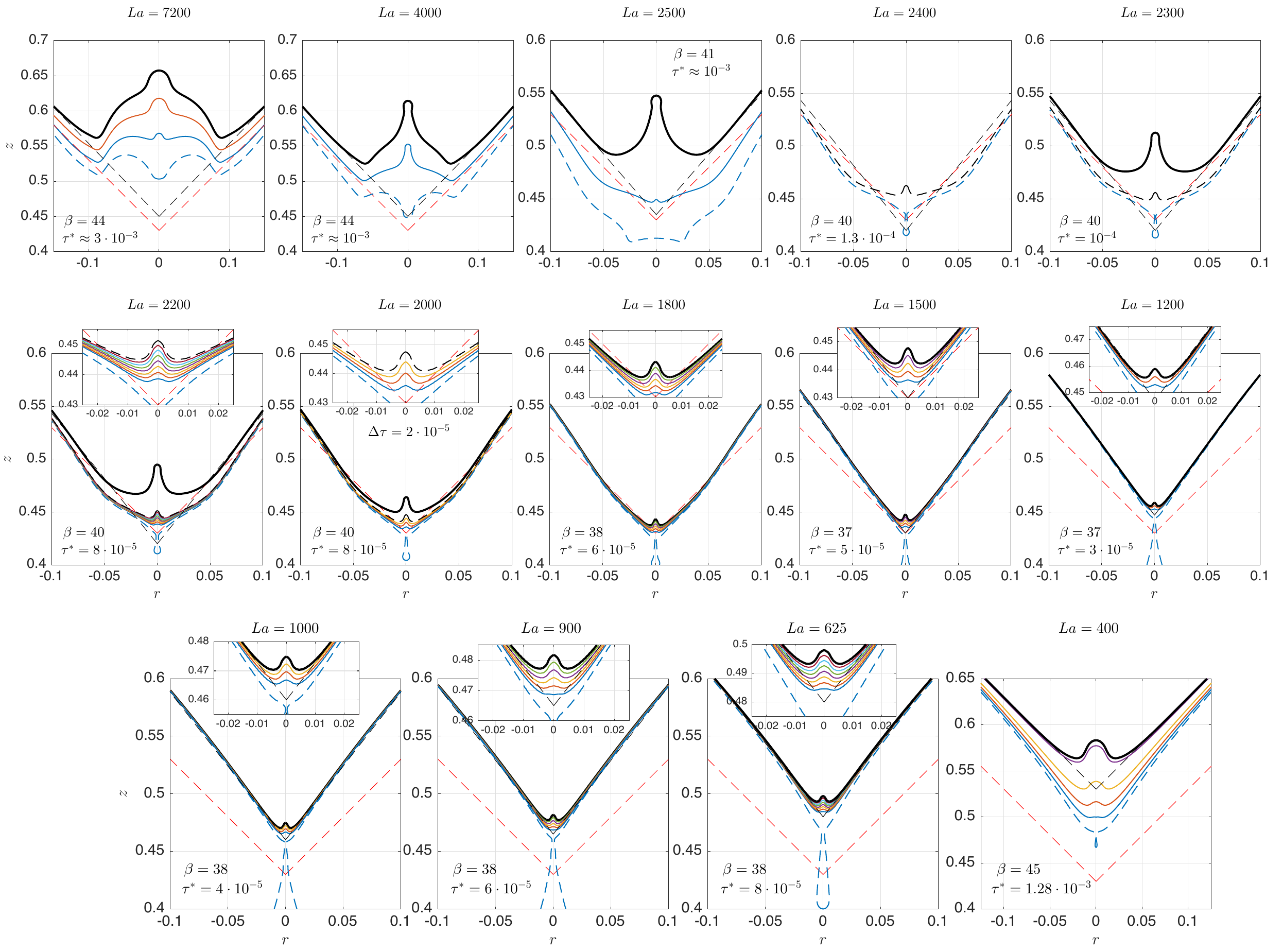}
	\caption{Initial time evolution of the cavity and the jet for different values of $La$.} 
   \label{fig3SM}
\end{figure}

Figure \ref{fig3SM} shows the very initial shapes of the jets for different values of $La$. The corresponding values of $\beta$ are indicated in each of the panels.

\subsection{Self-similar jet shapes}

\begin{figure*}
	\centering
	\includegraphics[width=\textwidth]{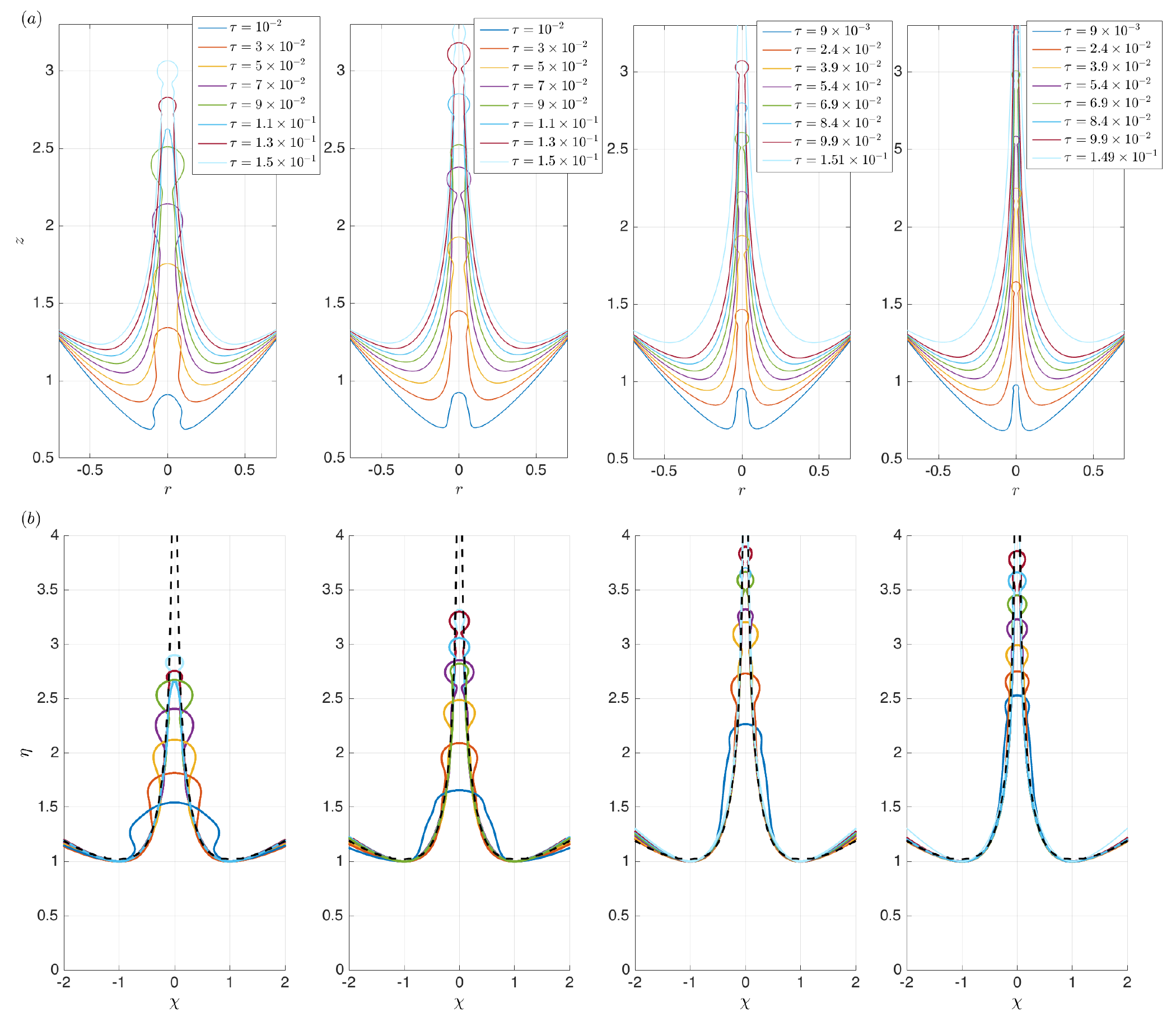}
	\caption{The jet shapes in the top row corresponding, from left to right to $La=17000$, $La=7200$, $La=4000$ and $La=2500$, overlap onto the universal self-similar and purely inertial solution depicted in Fig. 10 of \cite{PRF2023}, represented here in dashed lines, when plotted in terms of the scaled radial and vertical coordinates defined as $\chi=r/r_{jet}(\tau)$ and $\eta=(z-\ell_0)/(z_{jet}(\tau)-\ell_0)$, with $z-\ell_0$ representing the distance measured from the vertex of the cone with an opening semiangle $\beta\simeq 45^\circ$, see Figs. 1 and 5 in \cite{PRF2023}.} 
   \label{fig4SM}
\end{figure*}

The self-similar solution for the jet shapes reported in \cite{PRF2023} is compared in Fig. \ref{fig4SM} with the numerical results obtained in the main text for values of the Laplace number spanning from $La=17000$ to $La=2500$. 

\subsection{Values of $\tau^*(La)$}

\begin{figure}
	\centering
	\includegraphics[width=0.8\textwidth]{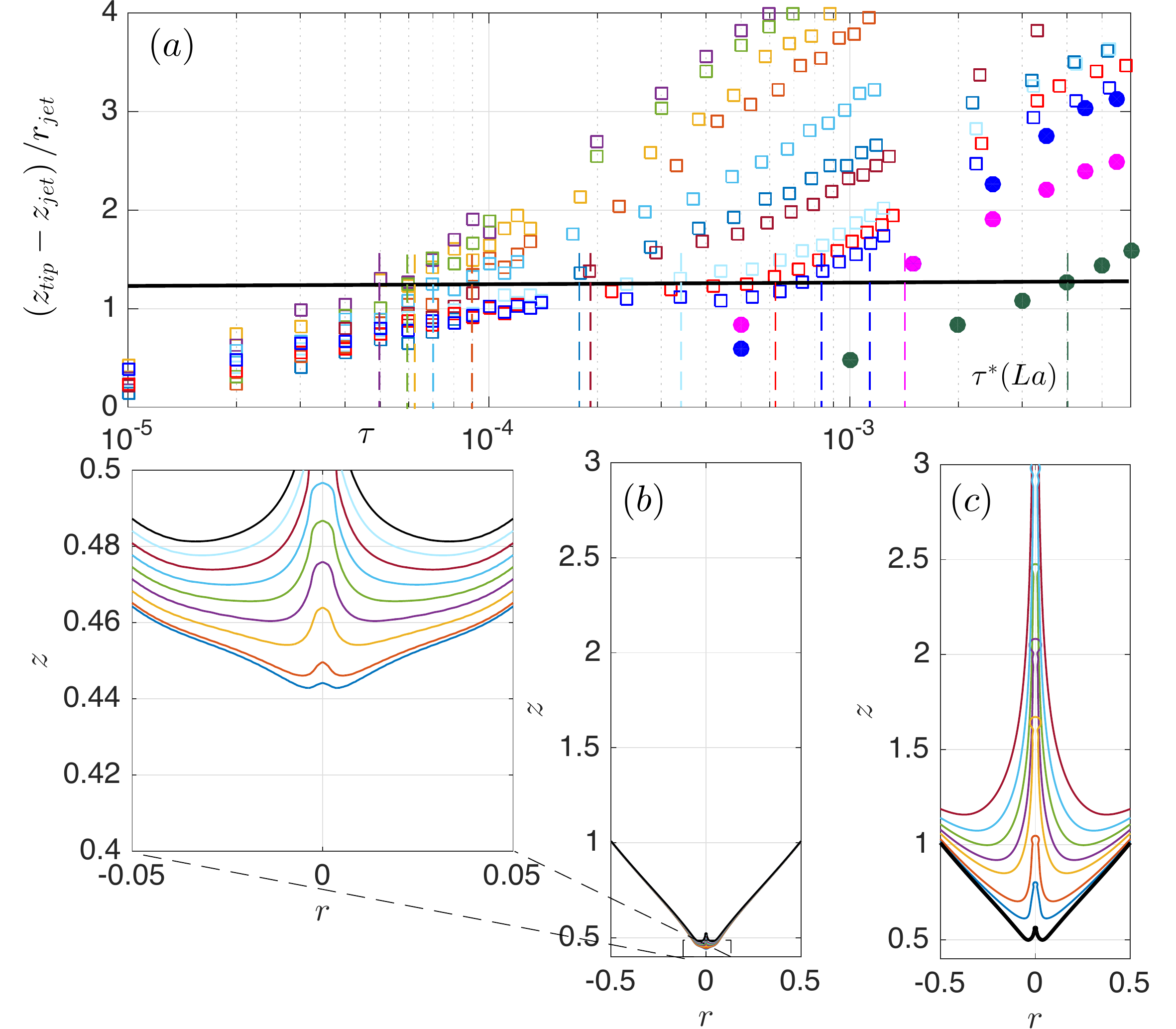}
	\caption{(a) The jet is issued when $(z_{tip}-z_{jet})/r_{jet}\gtrsim 1.25$: indeed, if $(z_{tip}-z_{jet})/r_{jet}<1.25$, the base of the cavity and the tip of the jet possess nearly the same velocity and, hence, a long and thin jet is not issued from the bulk of the liquid into the atmosphere, as it is shown for the case $La=2400$ in panel $(b)$. It can be seen in $(b)$ that a protuberance is formed at the base of the cavity, see the inset, but a jet with a length substantially larger than its width, is not produced when $\tau>0$ and $(z_{tip}-z_{jet})/r_{jet}<1.25$. In contrast, panel $(c)$ shows that a slender jet is clearly ejected for instants of times $\tau>0$ when $(z_{tip}-z_{jet})/r_{jet}\geq 1.25$. The vertical dashed lines indicate the values of $\tau^*(La)$.} 
   \label{fig5SM}
\end{figure}

Clearly, a necessary condition for a jet to be produced is $\tau=t-t_{bubble}>0$, but this condition is not sufficient: indeed, Fig. \ref{fig5SM}(b) shows that, if $\tau>0$ but $\left(z_{tip}-z_{jet}\right)/r_{jet}\lesssim 1$, with $r_{jet}(\tau)$ the jet width and $z_{tip}(\tau)$, $z_{jet}(\tau)$ respectively indicating the vertical positions of the tip and of the base of the jet -see the rightmost panel in Fig. 1(a) of the main text, a tiny protuberance appears at the bottom of the collapsing cavity but, in this case, the liquid bump is not clearly issued into the gaseous atmosphere because its aspect ratio is smaller than unity. In contrast, the results in Fig. \ref{fig5SM}(c) reveals that a long thin jet is indeed generated for all values of $La$ when $\tau>0$ and $\left(z_{tip}-z_{jet}\right)/r_{jet}\geq 1.25$, this being the reason why in this contribution we focus on the description of the time evolution of the jets produced for $\tau>\tau^*(La)$, with $\tau^*(La)>0$ the instant of time calculated from Fig. \ref{fig5SM}(a).

\subsection{Analysis of the collapse of the cavity before the jet is ejected}

\begin{figure*}
	\centering
	\includegraphics[width=0.9\textwidth]{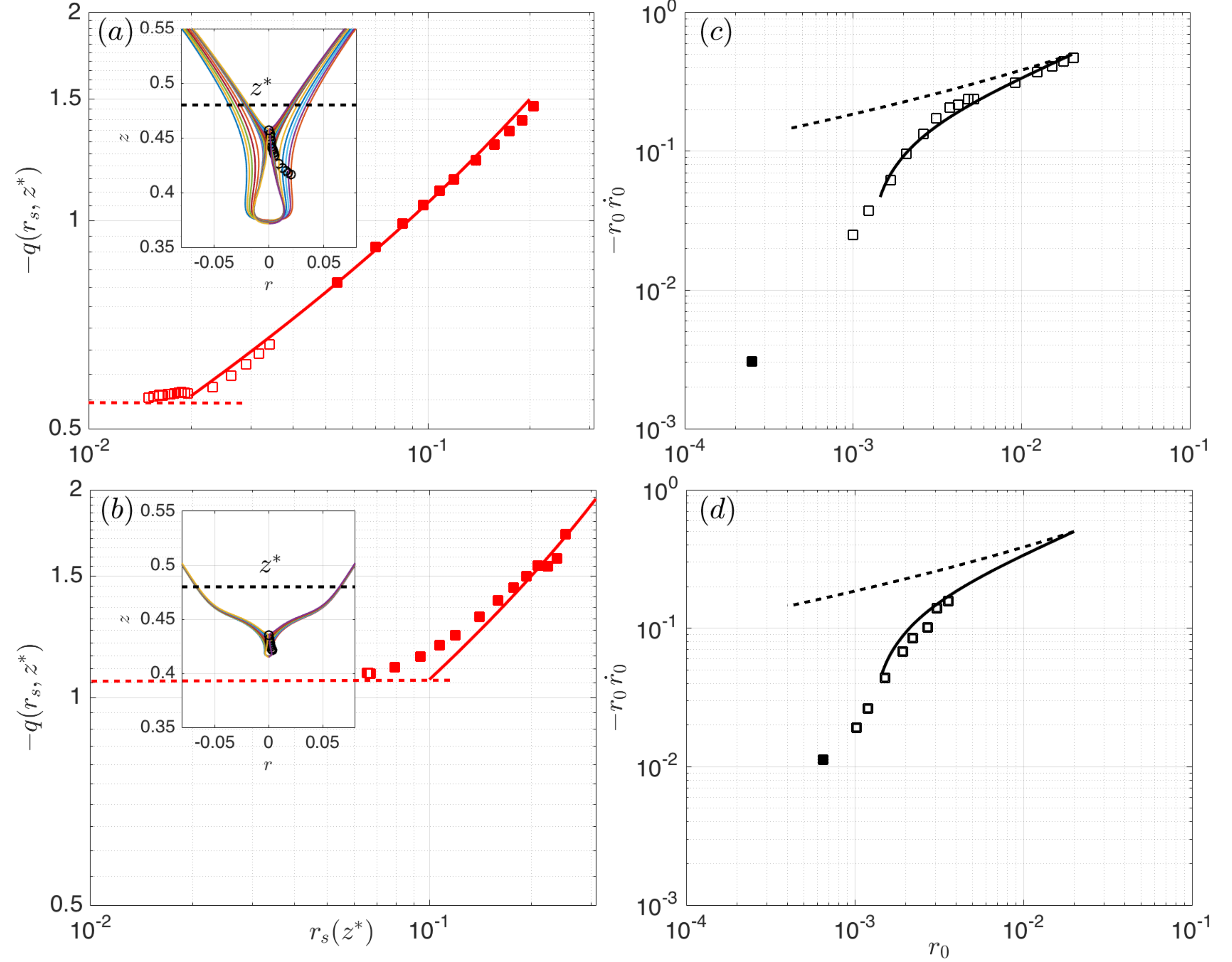}
	\caption{Numerical values of $q=-r_s(z^*,t) v_r(z*,t,r=r_s)$ for $La=1000$ $(a)$ and $La=2400$ $(b)$, calculated using \texttt{GERRIS}. Here, $r=r_s$ indicates the radial position of the interface and $v_r$ is the radial component of the velocity. The continuous red line represents the analytical solution of $q$ given in Eq. (\ref{Analitico}) corresponding to $r_0(0)=0.2$, $r_0\,\dot{r}_0(0)=-1.5$ and $r_0(0)/2\,r_c(0)=0.16$. The numerical time evolution of $q=-r_s(z_{min},t) v_r(z_{min},t,r_{min})$ with $r_{min}$ and $z_{min}$ indicating the radial and axial positions of the minimum radius of the cavity -indicated in panels ($a$) and ($b$) using hollow squares- for $La=1000$ (c) and $La=2400$ (d) are compared with the predicted values (plotted using continuous lines) obtained integrating the complete set of equations in Eq. (\ref{pinch}). The results reveal that both the influence of the gas and of the liquid viscosity need to be retained in the equations in order to reproduce the numerical results. Dashed lines correspond to the analytical solution  of $q$ given in Eq. (\ref{Analitico}). The initial conditions used to integrate the system (\ref{pinch}) in ($c$) and ($d$) are: $r_0=0.02$, $r_0\,\dot{r}_0(0)=-25$ and $r_0(0)/2\,r_c(0)=0.08$.} 
   \label{fig6SM}
\end{figure*}

This section is devoted to quantify the flow rate per unit length $q_\infty$ driving the ejection of the jet for $La<2500$. Within this range of $La$, a tiny satellite bubble is entrapped beneath the original bursting bubble before the jet is ejected, see Fig. \ref{fig2SM} and, motivated by this fact, the value of $q_\infty$ is calculated here as a function of the flow rate $q(t)$ driving the collapse of the walls of a nearly cylindrical bubble \cite{PRLBurton2005}. Indeed, here we calculate $q_\infty$ as $q_\infty=q(t^*)$ namely, as the value of $q(t)$ at the instant $t^*$ the jet is issued from the vertical position $z=z^*$ where the jet is firstly issued according to the criterion illustrated in Fig. \ref{fig5SM}. Figure \ref{fig6SM} illustrates the time-dependent values of $q$ calculated numerically using \texttt{GERRIS} as $q=-r_s(z^*,t) v_r(z^*,t,r=r_s)$ for instants of time before the jet is issued. Figure \ref{fig6SM} shows that $r_s(t,z^*,La=1000)<r_s(t,z^*,La=2400)$ and also that $q(t,z^*,La=1000)<q(t,z^*,La=2400)$. Figure \ref{fig6SM} also shows that $q$ decreases in time from values $q\simeq 1$ to values $q\simeq 0.5$ for the case $La=1000$ and also that $q$ remains mostly constant when the jet is about to be issued. Then, the results in Fig. \ref{fig6SM} indicate that the smaller is the radius of the cavity from which the jet is ejected, the smaller is $q(t)$ and, hence, according to our approximation $q_\infty=q(t^*)$, the smaller is the value of the flow rate per unit length $q_\infty$ driving the jet ejection process. Below, we explain the way the numerical values of $q(t)$ depicted in \ref{fig6SM} can be predicted within the range of values of the Laplace number $625\leq La\leq 2500$ for which a tiny satellite bubble is entrapped beneath the bursting bubble, see Fig. \ref{fig2SM}.

Indeed, in order to predict the value of the flow rate per unit length $q_\infty$ driving the jet ejection process, we need to  quantify first $q(t)$ as a function of the radius $r_0(t)$ of a nearly cylindrical collapsing bubble and, for this purpose, we make use here of the well-known equations describing the pinch-off of bubbles, see \cite{PRLBurton2005,PRL2005,PRLGiant,JFM2006,PRL07,PRL2007,PoF2008a}. In \cite{JFM2006,PoF2008a}, we analyzed the collapse of locally a parabolic axisymmetric cavity of equation
\begin{equation}
f(z,t)=r_0(t)+r_1(t)\,z^2=r_0(t)+\frac{z^2}{2\,r_c(t)}\label{f}
\end{equation}
with $r_0(t)$ and $r_c(t)$ respectively indicating the minimum radius of the cavity and its radius of curvature at $z=0$: indeed, when the gas flow rate through the neck is zero, it was shown in \citep{PRL2005,PRLGiant} that the cavity is symmetric around $z=0$ and also that the cavity is also locally slender i.e., $r_0(t)/(2r_c(t))\ll 1$. Defining $s=-\ln(r_0)$ and neglecting the subdominant capillary terms \cite{PRLBurton2005}, the leading order equations for $r_0(t)$ and $r_c(t)$ deduced in \cite{JFM2006,PoF2008a} describing the collapse of the cavity walls \emph{before the jet is issued} read
\begin{equation}
\begin{split}
&\ln\left(\frac{r_0}{2r_c}\right)\frac{d\ln \bar{q}}{ds}-1+\gamma\frac{\rho_g}{\rho}\frac{p}{\left[r_0/(2r_c)\right]}+\frac{4\,Oh}{\bar{q}\,r_0\dot{r}_0(0)}=0\, \\ &
\ln\left(\frac{r_0}{2r_c}\right)\frac{d}{ds}\left[\ln\left(\frac{r_0}{2r_c}\right)\right]-1+\frac{1}{2}\frac{\rho_g}{\rho}\frac{p}{\left[r_0/(2r_c)\right]}+\frac{4\,Oh}{\bar{q}\,r_0\dot{r}_0(0)}=0\, ,\\ & \mathrm{and}\quad\frac{d\,t}{ds}=-\frac{e^{-2s}}{r_0\dot{r}_0} \, ,
\label{pinch}
\end{split}
\end{equation}
with
\begin{equation}
\bar{q}=\frac{r_0\dot{r}_0}{r_0\dot{r}_0(0)}\label{Caudal}
\end{equation}
and $r_0\dot{r}_0(0)$ indicating the initial value of the product $r_0\dot{r}_0$. In \citep{PoF2008a} we found that $\gamma$ in Eq. (\ref{pinch}) is an order unity constant whose value can be approximated to $\gamma=0.5$. Moreover, in \citep{PoF2008a} we also found that the gas pressure gradient, $p$ in Eq. (\ref{pinch}), admits a self-similar solution which can be approximated by

\begin{equation}
p\simeq -8+\frac{16\,Oh}{r_0\dot{r}_0\,\left(\rho_g/\rho\right)\,\left(\mu/\mu_g\right)}\, .\label{p}
\end{equation}

Since $\gamma=0.5$, the subtraction of the two former equations in (\ref{pinch}) reduce, in the limits in which  $Oh\rightarrow 0$, $\rho_g/\rho\rightarrow 0$ and $\mu_g/\mu\rightarrow 0$, to
\begin{equation}
\frac{d}{ds}\left[\ln\left(\frac{r_0}{2r_c}\frac{1}{\bar{q}}\right)\right]=0\Rightarrow \bar{q}=\frac{r_0/\left(2r_c\right)}{\left[r_0/\left(2r_c\right)(0)\right]}\, .\label{qr0rc}
\end{equation}
Then, in the limit in which liquid viscosity and gas effects are negligible, the second of the equations in (\ref{pinch}) can be readily integrated to give
\begin{equation}
\begin{split}
&\ln\left(\frac{r_0}{2r_c}\right)\frac{d}{ds}\left[\ln\left(\frac{r_0}{2r_c}\right)\right]-1=0\Rightarrow \frac{d}{ds}\left[\ln\left(\frac{r_0}{2r_c}\right)\right]^2-2=0\Rightarrow\\& \left[\ln\left(\frac{r_0}{2r_c}\right)\right]^2-\left[\ln\left(\frac{r_0(0)}{2r_c(0)}\right)\right]^2=2\left(s-s(0)\right)=-2\ln\left(\frac{r_0}{r_0(0)}\right)\label{pinch+}
\end{split}
\end{equation}
a result which, together with the one in Eq. (\ref{qr0rc}) yields, see \cite{PoF2008a}:
\begin{equation}
\begin{split}
& r_0=r_0(0)\,e^{\left(1/2\right)\left[\ln(r_0(0)/(2 r_c(0))\right]^2}\,e^{-\left(1/2\right)\left[\ln(r_0/(2 r_c)\right]^2}\\ &
q(t)=-r_0\dot{r}_0=\frac{-r_0\dot{r}_0(0)}{\left[r_0/\left(2r_c\right)(0)\right]}\times \exp\left(-\sqrt{\left[\ln\left(\frac{r_0(0)}{2r_c(0)}\right)\right]^2-\ln\left(\frac{r_0}{r_0(0)}\right)^2}\right)\, .\label{Analitico}
\end{split}
\end{equation}

The results in Fig. \ref{fig6SM} reveal that there exists a good agreement between the numerical values of $q(t)$ and the predictions given by Eq. (\ref{Analitico}) taking $r_0(0)=0.2$, $r_0\,\dot{r}_0(0)=-1.5$ and $r_0(0)/2\,r_c(0)=0.16$ for all values of $La<2500$, this fact motivating the approximation used in the main text:

\begin{equation}
\begin{split}
&\mathrm{If}\quad r_{jet}\leq 0.05\quad    q_\infty(r_{jet})=-\frac{r_0\dot{r}_0(0)}{\left[r_0/\left(2r_c\right)(0)\right]}\times \exp\left(-\sqrt{\left[\ln\left(\frac{r_0(0)}{2r_c(0)}\right)\right]^2-\ln\left(\frac{r_{jet}}{r_0(0)}\right)^2}\right)\, \\ &
\mathrm{If}\quad r_{jet}>0.05\quad q_{\infty}=0.82.\label{qinftyrjet}
    \end{split}
\end{equation}

The reason for the different expressions of $q_\infty$ in Eq. (\ref{qinftyrjet}) relies on the fact that whenever $r_{jet}<r_{jet0}(La=2500)\simeq 0.05$, a tiny satellite bubble is entrapped beneath the bursting bubble before the jet is ejected and the equations for $q(t)$ are the ones describing the collapse of a cylindrical void with a time-varying radius $r_0(t)\ll 1$ which, hence, is much smaller than that of the original bubble. Then, for $La<2500$ the initial value of $q_\infty$ is determined by the local flow driving the collapse of a nearly cylindrical bubble whose radius $r_0(t)$ is much smaller than the radius of the initial bursting bubble. In contrast, when $r_{jet}>r_{jet0}(La=2500)\simeq 0.05$, a bubble is not entrapped, the radius of the cavity from which the jet first emerges is larger and the flow rate $q_\infty$ driving the jet ejection process is the one induced by the capillary collapse of the cavity at the length scale of the original bubble. 

In \cite{PRF2023} we deduce the following equations for the time evolution of the jet velocity, $v_{jet}(\tau)$, and for the jet width $r_{jet}(\tau)$:
\begin{equation}
v_{jet}=\frac{1.5\ K(\beta)}{\tan\beta}\sqrt{\frac{q_\infty(r_{jet})}{\tau}}\qquad \mathrm{and} \qquad
r_{jet}\,v_{jet}\simeq 3.4\,q_\infty(r_{jet})\label{vjetrjet}
\end{equation}
where we have taken into account that bubble bursting jets emerge from a conical cavity with a value of the opening semiangle $\beta\approx 40^\circ$ and, therefore, $K(\beta)\approx 1.6$ in \cite{PRF2023}. Hence, the equations for $r_{jet}(\tau)$ and $v_{jet}(\tau)$ in the main text are deduced solving Eqs. (\ref{qinftyrjet})-(\ref{vjetrjet}) taking $\beta(r_{jet}>0.05)=45^\circ$ and $\beta(r_{jet}<0.05)=38^\circ$, see Fig. 1(c) of the main text.

Once $r_{jet}(\tau)$ is known, the value of the local Weber number is calculated in the main text as:
\begin{equation}
    We_l(r_{jet}(\tau))=v^2_{jet} r_{jet}=\frac{\left[3.4\,q_\infty(r_{jet}(\tau))\right]^2}{r_{jet}(\tau)}\, \label{Werjet}
\end{equation}
where we have made use of Eqs. (\ref{qinftyrjet})-(\ref{vjetrjet}). 

Finally, Fig. \ref{fig6SM}(c)-(d) compares, for $La=1000$ and $La=2400$, the numerical time evolution of 
\begin{equation}
    q=-r_s(z_{min},t) v_r(z_{min},t,r_{min})
\end{equation}
 with $r_{min}$ and $z_{min}$ indicating the radial and axial positions of the minimum radius of the cavity, with the predicted value obtained integrating the complete set of equations in Eq. (\ref{pinch}). The results in Fig. \ref{fig6SM}(c)-(d) reveal that both the influence of the gas density and viscosity and of the liquid viscosity need to be retained in the equations in order to reproduce the numerical results.

\subsection{Initial values of the jet width and of the jet velocity}

\begin{figure*}
	\centering
	\includegraphics[width=0.9\textwidth]{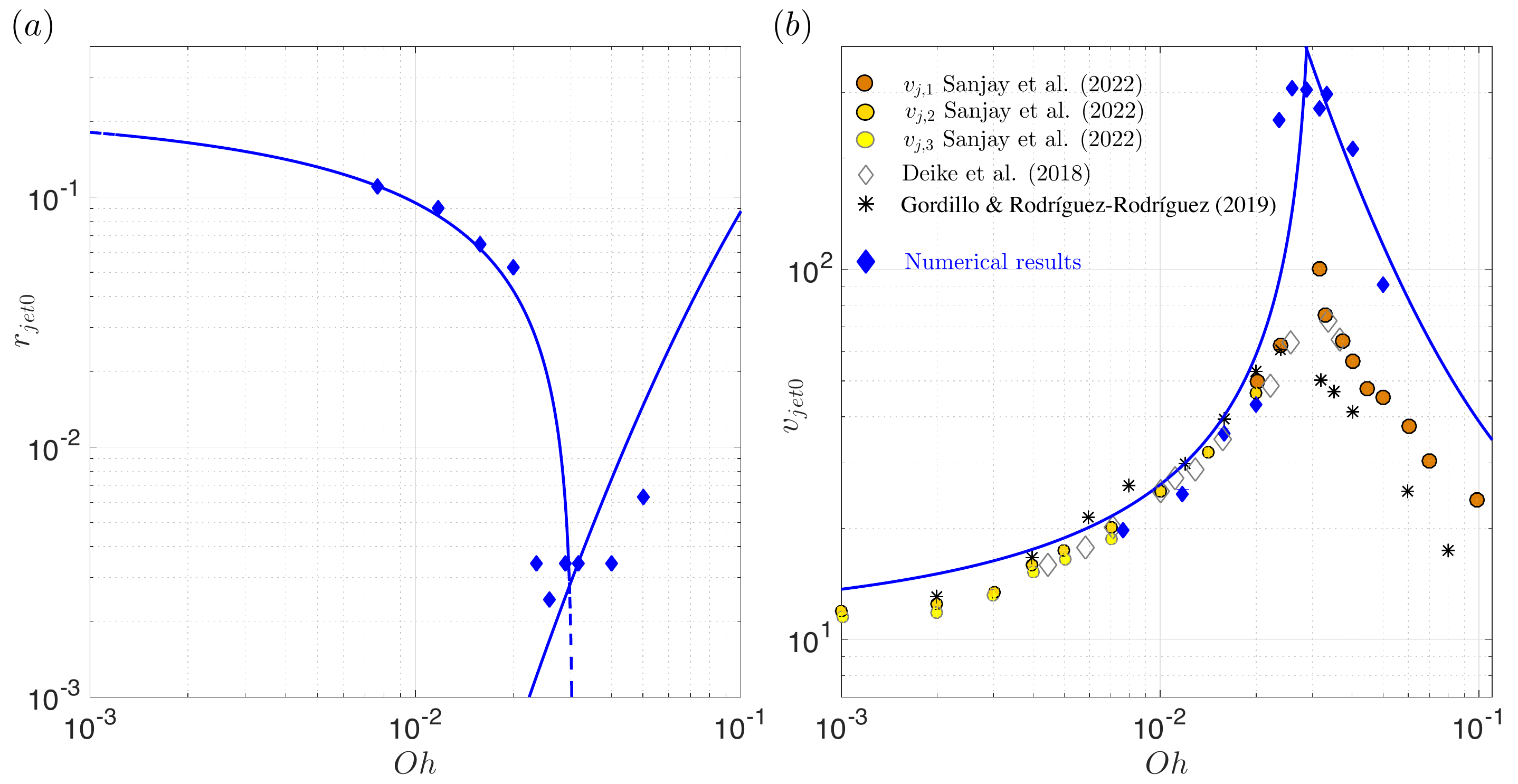}
	\caption{Comparison between the numerical values of $r_{jet0}$ and $v_{jet0}$ and the predictions using the equations deduced in the main text.} 
   \label{fig7SM}
\end{figure*}

Figure \ref{fig7SM} compares the predicted values for $r_{jet0}$ and $v_{jet0}$ given in the main text with the numerical values calculated here using \texttt{GERRIS} \cite{Popinet2009}. For $La>2500$, the numerical values correspond to the instants of time $\tau=\tau^*(La)$, see Fig. \ref{fig5SM}, whereas for $La\leq 1800$, the numerical values are those deduced from the analysis of the very initial instants of time depicted in Figs. \ref{fig3SM} and Fig. \ref{fig5SM} at $(z_{tip}-z_{jet})/r_{jet}\ll 1$. The numerical results in \cite{Sanjay2021} for $v_{jet0}$ are also included in Fig. \ref{fig7SM}(b). The differences existing between their results and ours are attributable to the differences existing in the instant of time at which the numerical values are calculated. In our case, the numerical values plotted in Fig. \ref{fig7SM} corresponds to instants of time $\tau\sim 10^{-5}$.

\bibliography{Gordillo_bibv2}
\end{document}